\documentclass[a4paper,11pt]{article}
\pdfoutput=1 

\usepackage{jheppub} 

\usepackage[T1]{fontenc} 
\usepackage{breqn}
\usepackage{derivative}
\usepackage{amsmath,mathrsfs}
\usepackage{subcaption}
\usepackage{appendix}
\usepackage[numbers,sort&compress]{natbib}
\usepackage{notoccite}

\title{\boldmath }

\title{Radiation from Axion star-Neutron star binaries with a tilted rotation axis in the presence of plasma}

\author[a]{A. Kyriazis}

\affiliation[a]{Department of Physics,\\ University of Florida , \\Gainesville, FL 32611, \\ United States }

\emailAdd{akyriazis@ufl.edu}

\abstract{We investigate the form of the radiation emitted by an axion star-neutron star binary using a $f(r)=\text{sech(r/R)}$ profile for the axion star. Our analysis takes into account the co-rotating plasma of the neutron star. We find that that there is significant enhancement to the radiated power if the neutron star's spin is tilted towards the plane of the axion star-neutron star orbit, compared to the case where it is perpendicular. We also examine whether the neutron star's co-rotating plasma can play a role in the emitted power and we find that even though dilute axion stars can in principle radiate more efficiently than dense axion stars, they will be pulled apart by the tidal forces of the neutron star}

\begin{document} 
\maketitle
\flushbottom

\section{Introduction}
\label{sec:intro}
\par The axion, initially proposed as a solution to the strong CP problem \cite{Peccei,Weinberg}, is now one of the most well-motivated candidates of dark matter \cite{Preskill,Sikivie, Willy}. The axions are stable bosons, with large occupation numbers and can rethermalize through their gravitational interactions forming a Bose-Einstein condensate (BEC) \cite{BEC1,BEC2}. Owing to the large occupation number of the ground state, the BEC condensate has been treated classically as a localised, coherently oscillating clump called an axion star, if the kinetic pressure is balanced by gravity and axiton or oscillon, if it is balanced by self-interactions. \cite{axion_star1,axion_star2,axion_star3,axion_star4,axion_star5,Chavanis,Wilczek}. 
\par The interaction of axions with electromagnetic field has been proposed as a venue for their detection \cite{axion_detection}. Of particular interest it the Primakoff effect \cite{Primakoff}, which is the interaction of an axion with a virtual photon to produce a real photon. 
One possible "laboratory" that exploits this effect for the detection of the axion is that of neutron stars, where some of the strongest magnetic fields in nature can be found \cite{axion_neutron}. 
\par The interaction between a dilute axion star (AS) and a neutron star (NS) has also been proposed to explain Fast Radio Bursts \cite{FRB1,FRB2,FRB3,FRB4}, where the magnetic field of the neutron star induces an electric field in the axion star which then accelerates electrons in the atmosphere of the neutron star \cite{FRB_axion_neutronstar_electron} or induces a dipole moment in neutrons on the upper upper core of the neutron star \cite{FRB_axion_neutronstar_neutron}. However, due to tidal forces, the AS will break apart in a time scale of the order of seconds, much longer than the millisecond duration of FRBs \cite{tidal}. It has been pointed out though that the axion to photon conversion in the collision of an \textit{dense} AS with a NS may still be able to explain FRBs \cite{FRB_dense,Bai}. 

\par Recently, the electromagnetic power emitted by an AS-NS binary system was derived analytically \cite{ANS_binaries}. It was assumed that the AS is described by an exponential profile, that the axis of rotation of the NS is perpendicular to the AS-NS plane and that plasma effects were negligible due to the low frequency of the plasma. For circular orbits, it was found that the spectral flux density of the emitted radiation is modulated mainly by the high frequency spin of the NS. There is also significant enhancement of emitted radiation for elliptical orbits due to the periodic approach of the AS to the NS.   
\par In this work, we include the plasma co-rotating with the NS through the equation $k=\sqrt{\omega^{2}-\omega^{2}_{p}}$, where k is the wave-number of the photon, $\omega$ is its frequency and $\omega_{p}$ is the plasma frequency. We also assume that the rotation axis of the neutron star has a tilt with respect to the the normal to the NS-AS plane, use a profile for the AS that is more accurate than the exponential in the case of attractive self-interactions \cite{Eby} and allows us to compare with previous works where a constant magnetic field was assumed \cite{dipole}, and finally explore the relevance of plasma effects when we consider an axion mass $m_{a} \sim \omega_{p}$.  
\par The paper is organised as follows: In section 2, we review the interactions between the axion and electromagnetism. In section 3, we establish the profile of the axion star that we will be using and describe the magnetic field of the pulsar. In section 4, we calculate the power emitted by the axion condensate and provide analytic formulae. In section 5, we summarise the results of this paper with some remarks for future research.

\section{Axion Electrodynamics}
The Lagrangian of the axion field coupled to electromagnetism is:
\begin{equation}
    \label{eqn:Lagrangian}
    \mathcal{L}=\frac{1}{2}\partial_{\mu}\partial^{\mu}\varphi-\frac{1}{2}m^{2}_{\varphi}\varphi^{2}-\frac{1}{4} F_{\mu \nu}F^{\mu \nu}-\frac{g_{a\gamma}}{4}\varphi F_{\mu \nu}\tilde{F}^{\mu \nu}
\end{equation}
where $\phi$ is the scalar field of the axion and:
\begin{equation}
    \label{eqn:Tensor}
    F_{\mu \nu}=\partial_{\mu}A_{\nu}-\partial_{\nu}A_{\mu} \,,
    \qquad
    \Tilde{F^{\mu \nu}}=\frac{1}{2}\epsilon^{\mu \nu \rho \sigma} F_{\rho \sigma}
\end{equation}
\par Note that in the above Lagrangian we have neglected gravity and assumed that there are no free currents and charges. Since the occupation number of the axion field is huge, we may treat its field as a classical field. The equations of motion are:
\begin{subequations}
\label{eqn: Maxwell}
\begin{align}
(\Box + m^{2}_{a}) \varphi = & g_{a\gamma} \bf{E} \cdot \bf{B}
\\
\nabla\cdot \bf{E} & =\rho 
\\
\nabla \times \bf{B} - \pdv{\bf{E}}{t} & = \bf{J}
\\
\nabla \cdot \bf{B} & =0
\\
\nabla \times \bf{E}+\pdv{\bf{B}}{t} & = 0
\end{align}

\end{subequations}
where the charge density $\rho$ and current density $\bf{J}$ are:

\begin{subequations}

\begin{align}
\rho  &= -g_{a\gamma}\nabla\varphi \cdot \textbf{B} \\
\textbf{J}  & = g_{a \gamma} (\pdv{\varphi}{t} \textbf{B}+\nabla \varphi \times \textbf{E}) \label{eqn: current} 
\end{align}
\end{subequations}

\par This is a set of coupled, partial differential equations. To make progress, we follow \cite{dipole} and expand the electromagnetic fields, charge density and current density in the small parameter $g_{a\gamma}\varphi_{0}$:

\begin{subequations}
\begin{align}
\bf{E}=\bf{E}^{(0)}+\bf{E}^{(1)}+...,
\qquad
\bf{B}=\bf{B}^{(0)}+\bf{B}^{(1)}+...
\end{align}
\end{subequations}
where $\bf{E}^{(0)}$ and $\bf{B}^{(0)}$ are the background fields that mix with the axion field. The 2 dynamical Maxwell equations for the first order electromagnetic fields are:

\begin{subequations}
\begin{align}
\nabla \cdot \textbf{E}^{(1)} & = \rho^{(1)},
\\
\nabla \times \textbf{B}^{(1)}-\pdv{\textbf{E}^{(1)}}{t} & = \textbf{J}^{(1)}
\end{align}
\end{subequations}

\par Introducing now the vector and scalar potential through the standard equations $\textbf{B}^{(1)}=\nabla \times \textbf{A}$, $\textbf{E}^{(1)}=-\nabla A^{0} - \pdv{\bf{A}}{t}$, our problem is reduced to calculating $A^{0}$ and $\textbf{A}$ from the equations:
\begin{subequations}
\begin{equation}
\label{eqn:wave equation potential}
(-\nabla^{2} + \frac{\partial^{2}}{\partial t^{2}} )A^{0} = \rho^{(1)} \\
\end{equation}
\begin{equation}
\label{eqn:wave equation vector}
(-\nabla^{2} + \frac{\partial^{2}}{ \partial t^{2}} )\textbf{A}  = \textbf{J}^{(1)}
\end{equation}
\end{subequations}
\par We now turn to the field of the AS and the background fields $\textbf{E}^{(0)}$ and $\textbf{B}^{(0)}$ of the NS.

\section{Axions stars and Neutron stars}
\subsection{Axion stars}

\par For the axion star, we use the Ansatz proposed in \cite{Schiappacasse_2018} for a spherical symmetric star, oscillating coherently with frequency $\omega$:
\begin{equation}
    \label{eqn:profile 1}
    \varphi=\varphi_{0}f(r) cos(\omega t)
\end{equation}
\par For the non-relativistic case that we examine here, we may set $\omega= m_{a}$. 
The spatial profile $f(r)$ must satisfy the boundary condition $ f(r) \xrightarrow[]{} 0$ when $r \xrightarrow[]{} \infty$ and have an inverted parabola at $r=0$. We will consider the  profile:
\begin{equation}
\label{eqn:profile 2}
    f(r) = \text{sech}(r/R)
\end{equation}
\par Although we should include a polynomial that captures the behaviour of the field at $r \xrightarrow[]{} 0$, these profiles give us the correct qualitative behaviour of the field. 
\\
It was found in \cite{Wilczek}, where the potential considered was that of the QCD axion:
\begin{equation}
    \label{eqn:potential}
    V(\varphi)=\Lambda^{4} (1-cos(\varphi/f_{a}))
\end{equation}
that if $|\varphi_{0}| \ll f_{a}\, (\frac{10^{-6} eV}{m_{a}})$, where $f_{a}$ is the axion decay constant, we are in the diluted, stable branch where $R \omega \gg 1$ and we can ignore the self-interactions. In this regime, the axion star is supported by its kinetic pressure acting against gravity. Its typical mass and radius is \cite{Chavanis,Wilczek}:
\begin{align}
\label{eqn: dilute AS}
    M_{dilute} & \leq 1.0 \times 10^{-9} M_{\odot} \left( \frac{f_{a}}{10^{13} \,\text{GeV}} \right) \left( \frac{10^{-6}\, \text{eV}}{m_{a}} \right) \\
    R_{dilute} & = 27 km \left( \frac{10^{13} \, \text{GeV}}{f_{a}} \right) \left( \frac{10^{-6} \, \text{eV} }{m_{a}} \right) 
\end{align}

If $|\varphi_{0}|=\mathcal{O}(1) f_{a} $, we are in the dense branch where $R \geq$few$   \times   \omega^{-1}$ and we have to include the self-interactions, while the non-relativistic approximation breaks down and our Ansatz \ref{eqn:profile 1} does not capture the correct profile of the oscillon. In addition, its lifetime was determined to be short, of the order $ \mathcal{O}(\frac{10^{3}}{m})$ and therefore not relevant to cosmological investigations. The radius of a dense axion star is given by \cite{axion_star1}:
\begin{equation}
\label{eqn: dense AS}
R_{dense} = 3 m \times \left( \frac{10^{13} \, \text{GeV}}{f_{a}} \right)^{1/2} \left( \frac{10^{-6} \, \text{eV}}{m_{a}} \right)^{1/2} \left( \frac{M_{dense}}{10^{-12} M_{\odot}} \right)^{0.3}
\end{equation}
\par In \cite{decay, Oscilon_dm,Structure_of_oscillon}, in addition to \ref{eqn:potential}, a number of other potentials were studied and the lifetime of oscillons was found to be up to 5 orders of magnitude larger. However, they arrived to similar results regarding the single frequency approximation i.e. we need a more sophisticated profile than \ref{eqn:profile 1} to study dense oscillons-axion stars. In this work, we will assume that the profile in \ref{eqn:profile 1} is also correct for dense axion stars and leave a multiple frequency profile for future work.

\subsection{Neutron Stars}
\par We are modelling the neutron star as a rotating magnetic dipole of angular frequency $\Omega$ surrounded by co-rotating plasma. In the near zone $r \ll \frac{1}{\Omega}$ the magnetic field is given by:
\begin{equation}
    \textbf{B}_{NS}=\frac{B_{0}}{2} \left(\frac{R_{NS}}{r} \right)^{3} (3(\hat{m} \cdot \hat{r}) \hat{r} - \hat{m})
\end{equation}
where $R_{NS}$ is the neutron star's radius, $B_{0}$ is the intensity of the magnetic field on the neutron star's surface and $\hat{m}$ is the unit vector in the direction of the magnetic moment. The number density of the electrons was derived by Goldreich and Julian assuming perfect conductivity of the plasma \cite{Goldreich}:

\begin{equation}
n_{c}=\frac{2 \bf{\Omega} \cdot \textbf{B}}{e} \frac{1}{1-\Omega^{2} r^{2} \sin^{2} \theta}
\end{equation}
while the plasma frequency is given by:

\begin{equation}
    \omega_{p}=\sqrt{\frac{4 \pi \alpha n_{c}}{m_{e}}} \approx 10^{-5} \text{eV} \sqrt{\frac{\hat{\Omega} \cdot \textbf{B}}{10^{12} G} \frac{1 s}{T_{s}}}
    \label{eqn: plasma freq}
\end{equation}
where $T_{s}$ is the period of the NS.
\begin{figure}
    \centering
    \includegraphics{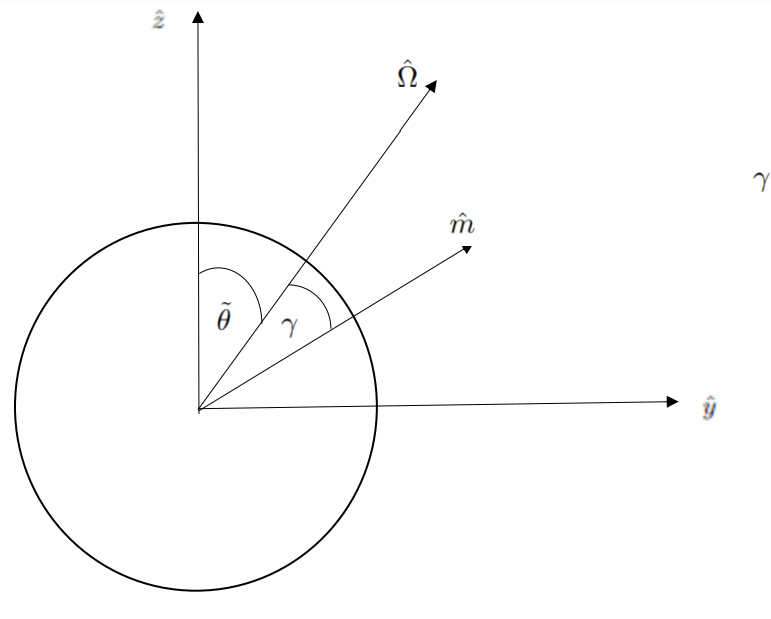}
    \caption{Schematic of the NS showing the different angles. The rotation axis makes a $\tilde{\theta}$ angle with the z axis and the magnetic dipole moment makes a constant $\gamma$ angle with the rotation axis, while precessing around it with $\Omega$ angular frequency}
    \label{fig:xi=0}
\end{figure}
\par We now turn to the rotation axis of the NS. For simplicity, we will assume that it lies on the y-z plane, at an angle $\tilde{\theta}$ from the z axis:
\begin{equation}
\label{eqn: rotation axis}
\hat{\Omega}=
\begin{pmatrix}
    0 \\ \sin\tilde{\theta} \\ \cos \tilde{\theta} \\
\end{pmatrix}
\end{equation}
We can now perform two successive rotations using SO(3) matrices to obtain the magnetic moment unit vector $\hat{m}$ that precesses around the rotation axis. One is around an axis perpendicular to $\hat{\Omega}$ by an angle $\gamma$ and the next one is around $\hat{\Omega}$ by an angle $\Omega t$. The details are given in Appendix \ref{appendix A}. Here, we simply write down the final answer:
\begin{equation}
    \label{eqn:magnetic moment}
    \hat{m}=
     \begin{pmatrix}
    \cos(\Omega t) \sin \gamma \\ \cos \gamma \sin \tilde{\theta} + \sin \gamma \cos \tilde{\theta} \sin(\Omega t) \\ \cos \gamma \cos \tilde{\theta} -\sin \gamma \sin \tilde{\theta} \sin (\Omega t)
    \end{pmatrix}
\end{equation}
\par Furthermore, we argue that we can ignore the background electric field in \ref{eqn: current}. Its value at the surface of the star is $E_{0}$=$B_{0} \Omega a$ and in order to ignore the second term in the right hand side of \ref{eqn: current}, it must hold:
\begin{equation}
    \nabla \varphi \times \textbf{E} \ll \pdv{\varphi}{t} \textbf{B} \Rightarrow \frac{\textbf{E}_{0}}{R} \ll m_{a} \textbf{B}_{0} \Rightarrow{} \frac{\Omega R_{NS}}{c} \ll m_{a}R 
\end{equation}
\par For typical values of $R_{NS}$=10 km, $\Omega=10^{-3}-1 s$, this inequality is true for the range of $m_{a}R$ that we are interested in. 

\subsection{AS-NS binary}
Assuming that $M_{NS} \gg M_{AS}$, we can place the NS in the origin of our coordinate system and the AS will be orbiting around it. The AS's motion is described by the familiar equation for elliptical orbits:
\begin{equation}
    r_{a}=\frac{(1-e^{2})r_{0}}{1+e \cos \phi_{a}} ,\ r_{0}=\frac{l^{2}_{0}}{G M_{NS}(1-e^{2})}
\end{equation}
where $r_{0}$ is the semi-major axis, $l_{0}$ is the angular momentum per unit mass, $\phi_{a}$ is the azimuthal angle of the AS and $e$ is the eccentricity. We will restrict our analysis to the cases where $e=0$ and $0 < e < 1$, to circular and elliptical orbits respectively. Since the AS lies on the $x-y$ plane, its position vector is:
\begin{equation}
    \textbf{r}_{a}=r_{a}(\cos \phi_{a}, \sin \phi_{a},0)
\end{equation}
Later on, we will also need the time dependence of the angle $\phi_{a}$ which is given by the parametric equations \cite{Goldstein}:
\begin{equation}
    \omega t = \psi - e \sin \psi ,\hspace{0.2 cm} \cos \phi_{a} = \frac{\cos \psi - e}{1 - e \cos \psi}
\end{equation}
where $0 \leq \psi \leq 2 \pi$ and $\omega$ is the angular frequency of the AS's rotation around the NS. The period of rotation is $T_{AS}=\frac{2 \pi }{\omega}$, which we will assume to be 10s. 
\par. We will also assume for simplicity that Earth's position vector lies on the z axis.

\begin{table}
\begin{center}
\begin{tabular} {| c | c | c | c |}
\hline
$M_{NS} (M_{\odot})$ & $M_{dense} (M_{\odot})$& $R_{NS} (km)$ & $r_{0} (km)$  \\
\hline
1.4 & $10^{-12}$ & 15 & $7.8 \times 10^{3}$ \\
\hline \hline
$\gamma (rad)$ & $T_{AS} (s)$ & $T_{s} (s)$ & $m_{a} (\text{eV})$\\
\hline
0.3 & 10 & 1 & $10^{-6}$ \\ 
\hline \hline
$f_{a} (\text{GeV})$ & $c_{\gamma}$ & $r_{e} (kpc)$ & $B_{0} (G)$\\
\hline
$10^{14}$ & $10^{5}$ & 1 & $10^{12}$ \\
\hline
\end{tabular}
\caption{A table with the relevant constants of our system}
\label{tab: table 1 }
\end{center}
\end{table}

\par Table \ref{tab: table 1 } summarises the constants relevant to our system, which we take to be the same as in \cite{ANS_binaries}, except for $f_{a}$ which we take to be $f_{a}=10^{13} \text{GeV}$, so that we are consistent with the equation $m_{a} f_{a}= m_{\pi} f_{\pi} = (10^{8} \text{eV})^{2}$ that must be true for the QCD axion.

\section{Analytic calculation of emitted power}

\par The radiating fields are given by the known equations:
\begin{align}
    \label{eqn:radiating fields}
    \textbf{B} & =\nabla \times \textbf{A}, \\
    \textbf{E} & =i \omega \textbf{A}-\nabla A^{0}=\frac{i}{\omega} \left(\omega^{2} \textbf{A} - \textbf{k} (\textbf{k} \cdot \textbf{A}) \right)
\end{align}

where we have taken the Fourier transform in time and used the Lorentz gauge $\nabla \cdot \textbf{A}+\partial_{0}A^{0}=0$. Here, $|\textbf{k}|=\sqrt{m_{a}^{2}-\omega_{p}^{2}}$. It is therefore sufficient to solve \ref{eqn:wave equation vector}. Using the retarded Green's function, the solution is:

\begin{equation}
    \textbf{A} (\textbf{r}_{e})  =\frac{g_{a \gamma} \varphi_{0} \omega \textbf{B}_{NS}}{4 \pi r_{e}}\int d^{3}r' e^{ik |\textbf{r}_{e} -\textbf{r}'|}  \, f(|\textbf{r}'-\textbf{r}_{a}|)
    \label{eqn: vector integral}
\end{equation}
The magnetic field varies negligibly within the AS, so we can take it out of the integral. We will only write the final result of this integration here and leave the details for Appendix \ref{Appendix B}.

\begin{equation}
    \textbf{A} (\textbf{r}_{e}) \approx \frac{g_{a \gamma} \varphi_{0} \omega (\pi R)^{2} \textbf{B}_{NS}(\textbf{r}_{a}) }{4 k r_{e}}  \frac{\tanh(\pi k R/2)}{\cosh(\pi k R/2)} e^{i k|\textbf{r}_{e}-\textbf{r}_{a}|}
    \label{eqn: vector solution}
\end{equation}
The time averaged Poynting vector is given by:
\begin{equation}
    \label{eqn:Poynting}
     \langle \textbf{S} \rangle = \frac{1}{2} \Re(\textbf{E} \times \textbf{B}^{\star})=\frac{1}{2 \omega} \Re \left( \textbf{k}(\omega^{2}|\textbf{A}|^{2}-|\textbf{k} \cdot \textbf{A}|^{2}) - \omega^{2}_{p} (\textbf{k} \cdot \textbf{A}) \textbf{A}^{\star} \right)
\end{equation}

Inserting in \ref{eqn:Poynting} the expression in \ref{eqn: vector solution}, we find for the Poynting vector and the power per solid angle :
\begin{equation}
\label{eqn: poynting solution}
    \langle \textbf{S} \rangle  = \frac{\pi^4 \omega}{32 r^{2}_{e}} \left( \frac{g_{a\gamma}\varphi_{0} R^{2}}{k} \right)^{2} \left( \frac{\tanh(\pi k R/2)}{\cosh(\pi k R/2)}\right)^{2} \left(\textbf{k}(\omega^{2}|\textbf{B}_{NS}|^{2} - (\textbf{k} \cdot \textbf{B}_{NS})^{2}) - \omega^{2}_{p} \textbf{B}_{NS}(\textbf{k} \cdot \textbf{B}_{NS}) \right)
    \end{equation}

\begin{equation}
        \label{eqn: solid angle }
        \frac{dP}{d \Omega}  = r^{2}_{e} \hat{k} \cdot \langle \textbf{S} \rangle = \frac{\pi^{4} (g_{a \gamma} \varphi_{0} \omega ^{2} R^{2})^{2}}{32 k \omega} \left( \frac{\tanh (\pi k R /2)}{\cosh (\pi k R /2)} \right)^{2} |\textbf{B}_{NS}|^{2} \sin^{2} \beta
\end{equation}

where $\beta$ is the angle between the magnetic field at the AS's position and Earth's position vector from the AS. It comes as no surprise that this is the same expression as the one derived in \cite{dipole}, where a constant magnetic field was assumed.

\begin{figure}
    \label{fig: power}
    \centering
    \includegraphics{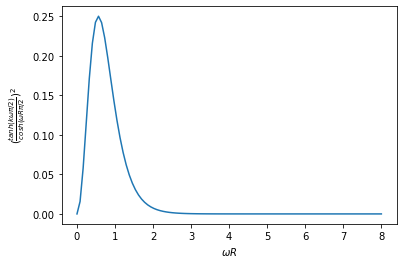}
    \caption{Plot of the function $\left( \frac{\tanh (\pi \omega R /2)}{\cosh (\pi \omega R /2)} \right)^{2}$ versus $\omega R$. The function has a peak at $\omega R$=0.56 and is exponentially suppressed for $\omega R \gg 1$}
\end{figure}
.

\subsection{AS-NS binary in vacuum}
\par For the benchmark values considered in Table \ref{tab: table 1 }, $\omega_{p} \sim 10^{-9} \text{eV}$, so we can neglect plasma effects for the moment, since $m_{a} \sim 10^{-6} \text{eV}$. The power per solid angle, equation \ref{eqn: solid angle }, is exponentially suppressed with respect to $\omega R$, which means that only dense axion stars can radiate efficiently in vacuum. We can understand this by considering energy-momentum conservation in the axion to photon conversion process \cite{Review}. 
\par
Let us examine the conversion rate of axions to photons in the presence of a magnetic field that has only spatial variation, $\textbf{B}_{0}(\textbf{x})$. The conversion rate is proportional to $| \int d^{3}x e^{-i (\textbf{k}-\textbf{k}_{a}) \cdot \textbf{x}} \hat{e}_{\lambda}(\hat{n})\cdot  \textbf{B}_{0}(\textbf{x}) |^{2}$, where $ \hat{e}_{\lambda}(\hat{n})$, with $\lambda=1,2$ are the two final state polarization vectors of a photon travelling in the $\hat{n}$ direction, $\textbf{k}$ is the momentum of the photon and $\textbf{k}_{a}$ is the momentum of the axion. A sum over $\lambda$ is implied. Energy conservation is satisfied because $\textbf{B}_{0}(\textbf{x})$ is time-independent. However, the axion satisfies the dispersion relation $\omega=\sqrt{k_{a}^{2}+m_{a}^{2}}$, while the photon satisfies $\omega=|\textbf{k}|$. Thus their momenta will in general be different, but if we decompose the magnetic field into its Fourier modes, $\tilde{\textbf{B}}_{0}(\textbf{q})$, momentum transfer is guaranteed because then $\textbf{q}=\textbf{k}-\textbf{k}_{a}$. In our case, we assumed that the magnetic field is constant within the range of integration, therefore $|\textbf{k}_{a}| \approx  \omega \Rightarrow \frac{1}{R} \approx \omega$. \par We may define the spectral flux density as $\mathcal{S}=\frac{dP}{d\Omega}\left(r^{2}_{e} \mathcal{B}\right)^{-1}$, where $\mathcal{B}=\frac{0.1 m_{a}}{2 \pi}$ is an estimate for the Doppler width. Inserting the benchmark values we are considering, the result is:
\begin{align}
\label{eqn: spectral dense}
    S & =  2.36 \times 10^{-5} \text{Jy} \left( \frac{c_{
    \gamma}}{10^{5}}  \right)^{2} \left( \frac{10^{13} \text{GeV}}{f_{a}} \right)^{2} \left( \frac{10^{-6} \text{eV}}{m_{a}} \right) \left( \frac{M_{dense}}{10^{-12} M_{\odot}}\right)^{1.2} \left( \frac{B_{0}}{10^{12} \text{GeV}} \right)^{2} \times \nonumber \\ & \times \left( \frac{1 kpc}{r_{e}} \right)^{2}  \sin^{2} \beta \left( \frac{1+e\cos \phi_{a}}{1-e^{2}} \right)^{6} |3(\hat{m} \cdot \hat{r})\hat{r} - \hat{m}|^{2}
\end{align}

\begin{figure}[h]
    \centering
    \includegraphics{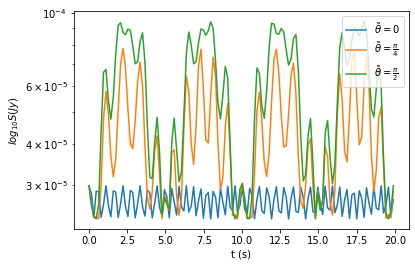}
    \caption{Spectral flux density versus time for e=0. The blue colour represents the rotation axis on the z axis, while orange and green represent the rotation axis at $\tilde{\theta}= \frac{\pi}{4}$ and $\tilde{\theta}= \frac{\pi}{2}$ respectively}
     \label{fig: e=0}
\end{figure}

In Figure \ref{fig: e=0}, we plot the spectral flux density versus time for three different $\tilde{\theta}$ angles. For $\tilde{\theta}=0$, our results are similar to what was found in \cite{ANS_binaries} i.e., the signal is modulated by the rotation of the AS around the NS and by the spinning of the NS around its axis. The envelope at the top is approximately flat because the separation between the AS and NS does not change with time. This picture changes if we allow the NS's rotation axis to be tilted. The position of the AS now plays a more important role in modulating the signal, as we can see from the peaks and troughs in the plot that have a period of 5 seconds. The locations of these maxima and minima can be explained if we remember that the AS starts its orbit on the x axis, where it receives the minimum amount of magnetic field lines from the NS, since the latter's rotation axis is now on the y-z plane. At 2.5 seconds, the AS has made one quarter of the orbit and it is found on the y-axis, where it now receives the maximum number of magnetic field lines. At 5 seconds, it is on the $x$ axis again, where it receives the minimum amount and so on. We would expect that $S$ would be larger the closer the NS rotation axis is to the x-y plane. This also confirmed by Figure \ref{fig: e=0}, where we see that as $\tilde{\theta}$ increases, $S$ also increases.

\begin{figure}[h]
    \centering
    \includegraphics{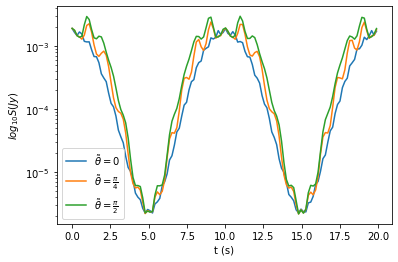}
    \caption{Spectral flux density versus time for e=0.5. The blue color represents the rotation axis on the z axis, while orange and green represent the rotation axis at $\tilde{\theta}= \frac{\pi}{4}$ and $\tilde{\theta}= \frac{\pi}{2}$ respectively}
   \label{fig: e=0.5}
\end{figure}

In the case where $e=0.5$, Figure \ref{fig: e=0.5}, the situation is different, because now the separation between the NS and the AS changes periodically with time. The orbit stars from the perihelion, where the signal is maximum. At 5 seconds, it is on the aphelion, where it is minimum and so on. As it was pointed out in \cite{ANS_binaries}, this is easily explained if we remember that $|\textbf{B}_{NS}|^{2} \sim \frac{1}{r^{6}}$, so there is significant enhancement to the signal if the AS is close to the NS. Hence, the modulation due to the NS spinning is not as important. This is confirmed by our results when we allow the NS's rotation axis to be tilted. In Figure \ref{fig: e=0.5}, it is evident that $S$ is slightly enhanced for the cases $\tilde{\theta}= \frac{\pi}{4}, \frac{\pi}{2}$, but not by much compared to the $\tilde{\theta}=0$ case. 

\subsection{AS-NS binary in plasma}
We would like to investigate how the plasma surrounding the NS affects the emitted power. To that end, we consider a lower axion mass, of the order $m_{a} = 10^{-9} - 10^{-8} \text{eV}$. This is outside of the QCD axion range, but it is valid for ALPs, where $f_{a}$ and $m_{a}$ are independent. We will keep the value of $f_{a}=10^{13} \text{GeV}$ for the axion decay constant.

\par It was shown in \cite{dipole} that when $\omega \rightarrow \omega_{p}$, the emitted power per solid angle becomes:
\begin{equation}
    \frac{dP}{d\Omega}({\omega \rightarrow \omega_{p}}) \approx  \frac{(g_{a \gamma} \varphi_{0})^{2}}{128} \frac{(\pi \omega R)^{6}}{\omega^{2}} \left(1-\frac{\omega_{p}^{2}}{\omega^{2}}\right)^{1/2} |\textbf{B}_{NS}|^{2} \sin^{2} \beta
\end{equation}
There is significant enhancement in power now if $\omega R \gg 1$,due to the factor $(\omega R)^{6}$. Therefore dilute axion stars can radiate much more efficiently than dense axion stars. Inserting in \ref{eqn: dilute AS} $m_{a}=10^{-9} \text{eV}$, we find for the radius $R_{AS,dilute}=27000 \text{km}$ and $M_{AS,dilute} \approx 10^{-6} M_{\odot}$. The Roche limit for this AS is then $R_{Roche}=R_{AS,dilute} \left(\frac{2 M_{NS}}{M_{AS.dilute}} \right)^{1/3} \approx 38 \times 10^{5} \text{km} $, much larger than the semi-major axis $r_{0}=7.8 \times 10^{3} \text{km}$. Hence, even though dilute axion stars could theoretically radiate in the presence of plasma, they will be pulled apart by tidal forces and disintegrate. We tried to avoid this problem by increasing $r_{0}$. However, if we increase $r_{0}$, the plasma frequency, equation \ref{eqn: plasma freq}, becomes smaller. This means that we have to lower the axion mass too, increasing the AS radius, via equation \ref{eqn: dilute AS}, leading again to a Roche limit larger than $r_{0}$.  It is an open question what happens to the remnants of this disintegrated AS which we leave for future work. 

\section{Conclusions and Outlook}
In this paper, we have derived analytical expressions for the spectral flux density emitted by an AS that is gravitationally bound to a NS. We have used a $\text{sech}$ profile to describe the AS that captures its essential characteristics and took into account plasma effects using the Goldreich-Julian model for the NS. We have also allowed the rotation axis of the NS to be tilted towards the plane of the AS-NS, placing it on the y-z plane. If the orbit is circular, we find that there is an order of magnitude increase to the spectral flux density whenever the AS is on the y axis, because of the increased number of magnetic field lines that now pass through the axion star. If, on the other hand, the orbit is an ellipse, there is no significant enhancement to the flux, since now the main modulation comes from the AS's orbit. 
\par
We also considered the effect that the plasma may have on the flux. Given that in the presence of plasma only dilute stars can radiate efficiently, we lowered the axion mass to $m_{a} = 10^{-9} \text{eV}$ to investigate this possibility. We found that by lowering the mass, we increase the AS radius, which in turn increases the Roche limit by several orders of magnitude more than the AS-NS distance, meaning that the dilute AS will be torn apart by tidal forces before it interacts with the NS's magnetic field. 
\par 
Some interesting venues for future research would be to investigate the effect of back-reaction to the AS profile. This would probably require solving numerically the set of equations \ref{eqn: Maxwell}. Back-reaction could also shed some light to the time scale of disintegration of a dilute AS in the gravitational field of a neutron star or a black hole. Another interesting direction of research would be to include a multi-frequency profile for the dense AS. As it was mentioned in the introduction, given that there have been attempts to explain FRBs with dense axion stars, a more accurate profile for the AS would be a step in that direction.

\acknowledgments
I would like to thank David Sadek, Jose Alberto Ruiz Cembranos, Dennis Maseizik and especially Pierre Sikivie for useful discussions. 

\appendix
\section{Appendix }
\label{appendix A}
We choose the vector perpendicular to $\hat{\Omega}$ to be: 
\begin{equation}
\hat{\Omega}_{\perp}=
    \begin{pmatrix}
        0 \\ -\cos \tilde{\theta} \\ \sin \tilde{\theta}
    \end{pmatrix}
\end{equation}
The rotation matrix about this vector and by an angle $\gamma$ is given by $R(\hat{\Omega}_{\perp},\gamma)=e^{i \gamma \, \hat{\Omega}_{\perp} \cdot \textbf{L}}$, where $\textbf{L}$ are the generators of SO(3). They are given by: 
\begin{equation}
\label{eqn:generators}
L_{1} =
\begin{pmatrix}
0 & 0 & 0 \\ 0 & 0 & -i \\ 0 & i & 0 \\
\end{pmatrix}
,\
L_{2} =
\begin{pmatrix}
0 & 0 & i \\ 0 & 0 & 0 \\ -i & 0 & 0 \\
\end{pmatrix}
,\
L_{3} =
\begin{pmatrix}
0 & -i & 0 \\ i & 0 & 0 \\ 0 & 0 & 0 \\
\end{pmatrix}
\end{equation}
Using these expressions, we find the rotation matrix to be:

\begin{equation}
   R(\hat{\Omega}_{\perp},\gamma)=
   \begin{pmatrix}
\cos \gamma & \sin \tilde{\theta} \sin \gamma & \cos \tilde{\theta} \sin \gamma \\ -\sin \tilde{\theta} \sin \gamma  & \cos^{2} \tilde{\theta} + \sin^{2} \tilde{\theta} \cos \gamma & -\sin 2 \tilde{\theta} \sin^{2} (\frac{\gamma}{2}) \\ -\cos \tilde{\theta} \sin \gamma & -\sin (2\tilde{\theta}) \sin^{2}(\frac{\gamma}{2}) & \sin^{2} \tilde{\theta}+\cos^{2} \tilde{\theta} \cos \gamma \\
\end{pmatrix}
\end{equation}
As it was mentioned in the text, we now perform a second rotation around $\hat{\Omega}$ by an angle $\Omega t$. The matrix is $R(\hat{\Omega},\Omega t)=e^{-i \, (\Omega t) \, \hat{\Omega} \cdot \textbf{L}}$ and using the generators \ref{eqn:generators}, we get:
\begin{equation}
   \label{eqn: first rotation matrix} 
   R(\hat{\Omega},\Omega t)=
   \begin{pmatrix}
\cos \Omega t & -\cos \tilde{\theta} \sin \Omega t & \sin \tilde{\theta} \sin \Omega t \\ \cos \tilde{\theta} \sin \Omega t  & \sin^{2} \tilde{\theta} + \cos^{2} \tilde{\theta} \cos \Omega t & \sin 2 \tilde{\theta} \sin^{2} (\frac{\Omega t}{2}) \\ -\sin \tilde{\theta} \sin \Omega t & \sin (2\tilde{\theta}) \sin^{2}(\frac{\Omega t}{2}) & \cos^{2} \tilde{\theta}+\sin^{2} \tilde{\theta} \cos \Omega t \\
\end{pmatrix}
\end{equation}
Applying these two rotations successively, we obtain the magnetic moment unit vector in equation \ref{eqn:magnetic moment}.

\section{Appendix}
\label{Appendix B}
The integral of equation \ref{eqn: vector integral} is:
\begin{equation}
\label{eqn: start int}
   \mathcal{I} = \int d^{3} r' e^{ik|\textbf{r}_{e}-\textbf{r}'|} f(|\textbf{r}'-\textbf{r}_{a}|)
\end{equation}
The easiest way to do it is to shift the coordinate system and place in the centre of the AS, $\textbf{r}' \rightarrow \textbf{r}_{a} + \textbf{s}$. The exponential can be written as $|\textbf{r}_{e}-\textbf{r}_{a}-\textbf{s}| \approx |\textbf{r}_{e}-\textbf{r}_{a}| - \frac{( \textbf{r}_{e}-\textbf{r}_{a}) \cdot \tilde{\textbf{r}}}{|\textbf{r}_{e}-\textbf{r}_{a}|} \approx |\textbf{r}_{e}-\textbf{r}_{a}| - \hat{r}_{e} \cdot \textbf{s} $, where we have assumed that $r_{e} \gg r_{a}$. Of course, $\omega_{p}(r') \sim \frac{1}{(r')^{3/2}} = \frac{1}{|\textbf{s}+\textbf{r}_{a}|^{3/2}}$. But, because of the exponentially decaying profile $\text{sech}(r/R)$ that we are using for the axion star, the main contribution to the integral \ref{eqn: start int} will come from $0 < s \lesssim 5 R$, much smaller than $r_{a}$. We therefore set $\omega_{p}(r')=\omega_{p}(r_{a})$.The angular part of the integral is:
\begin{equation}
    \label{eqn: angular}
    2 \pi \int^{\pi}_{0} d \theta_{s} \sin \theta_{s} e^{-i k s \cos \theta_{s}}=\frac{4 \pi \sin(ks)}{ks} 
\end{equation}
while, using the profile $f(r)=\text{sech}(r/R)$, the radial part becomes:
\begin{equation}
    \label{eqn: radial}
    \int^{\infty}_{0} ds \, s \, \sin(ks) \, \text{sech}(s/R)=\frac{\pi^{2} R^{2}}{4} \, \frac{\tanh \left( \frac{\pi k R}{2} \right)}{\cosh \left( \frac{k R \pi }{2} \right)}
\end{equation}
Plugging equations \ref{eqn: angular} and \ref{eqn: radial} back into \ref{eqn: start int}, we get:
\begin{equation}
    \mathcal{I} = \frac{\left( \pi R \right)^{3}}{k R} \frac{\tanh \left(\frac{k R \pi}{2} \right)}{\cosh \left( \frac{k R \pi}{2} \right)} e^{ik |\textbf{r}_{e} - \textbf{r}_{a}|}
\end{equation}
which gives equation \ref{eqn: vector solution}.
\bibliography{cit}

\end{document}